\documentclass[a4paper]{jpconf}
\usepackage{graphicx}
\begin{document}
\title{Selected topics on Hadrons in Nuclei}

\author{E. Oset$^1$, M. Kaskulov$^2$, E. Hernandez$^3$, A. Ramos$^4$, V.K. Magas$^4$, J. Yamagata-Sekihara$^1$, S. Hirenzaki$^5$, D. Gamermann$^4$, R. Molina$^1$, L. Tolos$^6$, L. Roca$^7$ }

\address{$^1$Departamento de F\'{\i}sica Te\'orica and IFIC,
Centro Mixto Universidad de Valencia-CSIC, \\
\small Institutos de
Investigaci\'on de Paterna, Aptd. 22085, 46071 Valencia, Spain \\
$^2$Institut f¨ur Theoretische Physik, Universitaet Giessen, D-35392 Giessen, Germany\\
 $^3$~Grupo de F\'{\i}sica Nuclear, Departamento de Fisica Fundamental
e IUFFyM, \\
\small Facultad de Ciencias, 
Universidad de Salamanca, \\
\small Plaza de la Merced s/n,
E-37008 Salamanca, Spain\\
$^4$Departament d'Estructura i Constituents de la Mat\`eria 
and Institut de Ci\`encies del Cosmos, 
Universitat de
Barcelona, 08028 Barcelona, Spain\\
$^5$Department of Physics, Nara Women's University, Nara 630-8506, Japan\\
$^6$Theory Group. KVI. University of Groningen,
Zernikelaan 25, 9747 AA Groningen, The Netherlands\\
$^7$Departamento de Fisica. Universidad de Murcia. E-30071, Murcia. Spain}
\ead{oset@ific.uv.es}

\begin{abstract}
In this talk we report on selected topics on hadrons in nuclei. The first topic is the renormalization of the width of the $\Lambda(1520)$ in a nuclear medium. This is followed by a short update of the situation of the $\omega$ in the medium.  The investigation of the properties of $\bar{K}$ in the nuclear medium from the study of the $(K_{flight},p)$ reaction is also addressed, as well as properties of X,Y,Z charmed and hidden charm resonances in a nuclear medium. Finally we address the novel issue of multimeson states. 
\end{abstract}

\section{Introduction}
  
 The last decade has experienced an important push in hadron spectroscopy 
 \cite{klempt,crede,Oller:2000ma,Oset:2009jd} and we have also come to realize that the old concept of mesons made out of $q \bar{q}$ and baryons made out of three $q$ is too simple and the reality is more complex in many cases. One of the fruitful ideas is that many mesonic and baryonic  resonances are dynamically generated from the interaction of more elementary hadrons. We call these states dynamically generated states, or molecular states, since they are obtained from the interaction of the hadronic components, much as a deuteron is a bound state of a proton and a neutron. The many predictions that one can do with the theory and the comparison with the data reinforce this assumption concerning the nature of these states in multiple cases. In the following we will address some of these states. The issue of the interaction of kaons in nuclei has also raised much interest lately and efforts have been devoted to find signatures of possible deeply bound states or to obtain the optical potential describing the interaction of $\bar{K}$ with nuclei. Understanding the vector meson properties in a medium has also received continuous attention in the latest years and we will  over only one topic in this issue, the $\omega$ in the medium focussing on the latest experimental results. With new facilities producing charmed and hidden charm states, the study of such systems in a nuclear medium has also raised expectations and we will address this problem here.  Finally, we devote a few lines to recall why nuclei are made of baryons and how far one can go producing aggregates of mesons rather than baryons.
 
 \section{The $\Lambda(1520)$ in the nuclear medium}   
  
\begin{figure}[t]
\begin{center}
\includegraphics[clip=true,width=0.45\columnwidth,angle=0.]
{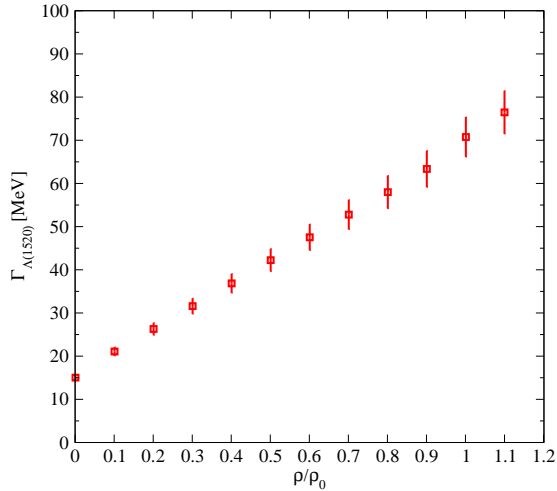}
\caption{ \label{F2} \small
The $\Lambda(1520)$ width in the nuclear medium 
as a function
of the density $\rho/\rho_0$ where $\rho_0$ 
is the normal nuclear matter density. }
\end{center}
\end{figure}

The $\Lambda(1520)$ $(J^P=3/2^-)$) is one of the resonances which is dynamically generated  from the interaction  of pseudoscalar mesons with the decuplet of baryons \cite{kolodecu,Sarkar:2004jh,Roca:2006sz}. The problem is initiated with the channels $\pi \Sigma(1385)$, $K \Xi^*$ in s-wave, but there are the $\bar{K} N$ and $\pi \Sigma$ states in d-waves which are also relevant and are the main decay channels of the resonance. The $\pi \Sigma(1385)$ appears as the main building block in the $\Lambda(1520)$, yet, this is not obvious experimentally since there is no decay of the resonance in this channel because this component is energetically closed. Nuclear physics gives us an opportunity to test this hypothesis. Indeed, if we are in the medium the $\pi$ can excite a $p-h$ (particle-hole) component and then the energy is a few MeV instead of the 140~MeV of the pion.  As a consequence of the strong coupling of the resonance to the $\pi \Sigma(1385)$ channel and the gain of 140~MeV of phase space for decay into 
$p-h \Sigma(1385)$, the width of the resonance is considerably increased. The problem was discussed in \cite{Kaskulov:2005uw} and we show the results in Fig.~\ref{F2}.

 As one can see in the figure there is a spectacular increase in the width of the resonance as the nuclear density increases going from 15 MeV in the free case to about 70 MeV at $\rho=\rho_0$.  Such a drastic change could be observed experimentally and in \cite{Kaskulov:2006nm} the transparency ratio for photon- and proton-induced $\Lambda(1520)$ production was studied. Clear effects were seen, which advise the performance of such experiments to test the strong coupling predicted of the resonance to the $\pi \Sigma(1385)$ channel, which would reinforce the concept of this resonance as a dynamically generated state.
 
 \section{The $\gamma A \to \omega X \to (\pi^0 \gamma) X$ reaction and the $\omega$ in the medium} 
 
   The modification of vector meson properties in the nuclear medium has attracted much attention, both theoretically and experimentally, in the last years and some recent reviews give account of this intense activity 
   \cite{Rapp:1999ej,Hayano:2008vn,Leupold:2009kz}. Here, we will briefly summarize recent developments concerning the $\omega$ in the medium.
   The TAPS/ELSA Collaboration reported in \cite{Trnka:2005ey} evidence of an attractive mass shift of the $\omega$ in the nuclear medium of about 200 MeV at normal nuclear matter density. This finding was immediately questioned in  \cite{Kaskulov:2006zc} where it was pointed out that the conclusions were tied to the choice of background made in \cite{Trnka:2005ey}, where a big change on  the shape of the background in nuclei with respect to the one on the proton was made. It was shown in \cite{Kaskulov:2006zc} that other reasonable assumptions for the background did not support that conclusion. Attempts were made in \cite{Metag:2007hq} to justify the background chosen in \cite{Trnka:2005ey} using the mixed event method to determine the background. Yet, this method was proved unsuited \cite{Kaskulov:2010za} since, irrelevant of the actual background, the procedure always provides the same background at energies in the  upper side of the $\omega$ mass distribution.
   
\begin{figure}[t]
\begin{center}
\includegraphics[clip=true,width=0.40\columnwidth,angle=0.]
{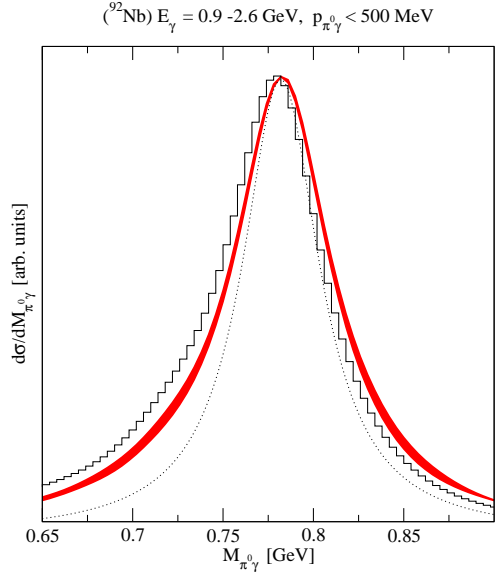}
\caption{\label{PRL2} \footnotesize
$\omega$ invariant mass spectra reconstructed from $\pi^0\gamma$ events.
The band corresponds to the changes of the in--medium $\omega$-mass
according to $m_{\omega}\pm 40 (\rho/\rho_0) $~MeV. The serrated line 
corresponds
to the scaling of the $\omega$ mass $m_{\omega}(1- 0.16 \rho/\rho_0)$.
The dotted curve is the $\omega\to \pi^0\gamma$ signal without
applying  kinematic cuts. 
}
\end{center}
\end{figure}

    It was also shown in \cite{Kaskulov:2006zc} that the reaction of \cite{Trnka:2005ey} was unsuited to determine the mass shift in the medium since the cross section barely changed within a variaton range in the mass of 80 MeV, as seen in Fig. \ref{PRL2}. 
    At some time, some peaks seen in the experiment looked like promising signals of possible $\omega$ bound states in nuclei, but they were also interpreted as a consequence of a different behavior of the background and the $\omega$ signal as a function of  the mass number \cite{Kaskulov:2006fi}. 
    
      The confusion on the subject led the experimental group to make an explicit determination of the background and the latest results have been disclosed only recently \cite{Nanova:2010sy}. A convergence between the theoretical findings and the experiment has been reached, and quoting textually from \cite{Nanova:2010sy} we read "The fact that the experimental data is consistent with both scenarios (one with no shift and the other with 125 MeV shift of $\omega$ mass) indicates that the line shape is indeed insensitive to in-medium modifications (of the $\omega$ mass) for the given invariant mass resolution and statistics. An earlier claim (\cite{Trnka:2005ey}) of an in-medium lowering of the $\omega$ mass is not confirmed."
      
         On the other hand, one should mention a more positive aspect of the work of the TAPS/ELSA collaboration where a spectacular increase  in the width of the $\omega$ in the medium has been found. Based on the theoretical studies of the transparency ratio in  \cite{Kaskulov:2006zc,Muhlich:2003tj} the measurements of the $\omega$ production cross section in \cite{:2008xy} revealed that the width of the $\omega$ in the medium at nuclear matter density is about 100 to 150 MeV, much larger than the 8.4 MeV width in vacuum.  Such spectacular increase is a challenge for theories trying to understand the medium modification of the vector meson properties.
	 
	 Regarding other recent theoretical developments on vector mesons in nuclei, a novel information comes from the study of the modification of the $\bar{K}^*$ in the medium, where an increase of the width by about a factor five in the nuclear medium is found \cite{Tolos:2010fq}. This work is presented by Laura Tolos in this same session. 
	 
\section{The $\bar{K}$ nucleus potential deduced from the in flight $(K^-,N)$ reaction}

  The issue of the $\bar{K}$ nucleus potential has also received much attention lately, with hopes that a sufficiently deep potential might lead to deeply bound kaon states. Recent updates of the experimentaÃl and theoretical situation can be seen in \cite{Oset:2009kf,Ramos:2008zza}. One has two extremes, one highly attractive phenomenological potential having a strength of
about 600 MeV at the center of the nucleus, leading to nuclear densities ten times that of normal nuclear matter \cite{Akaishi:2002bg}, and the more moderate potentials with a strength of about 50 MeV at normal nuclear matter density, which are obtained from chiral unitary approaches with selfconsistency 
\cite{Lutz:1997wt,Ramos:1999ku,SchaffnerBielich:1999cp,Cieply:2001yg,Tolos:2006ny}. All these potentials lead to deeply bound $K^-$ atoms, only that the chiral potentials produce states with a width too wide to lead to neat experimental peaks. While different experimental claims of deeply bound states have been ruled out, showing that the peaks observed have a conventional explanation in terms of theoretically controlled unavoidable reactions \cite{Oset:2009kf,Ramos:2008zza}, a recent experiment claimed to have evidence that the depth of the potential was of the order of 200 MeV attractive at normal nuclear matter density \cite{Kishimoto:2007zz}. The reaction is the in flight $(K^-,N)$ process in nuclei using a beam of 1 GeV/c $K^-$ against a nuclear target of $^{12}C$ and looking at energetic protons or neutrons (500-700 MeV of kinetic energy) in the forward direction. In these kinematics one must take into account that the kaons go backwards in the center of mass frame. From the shape of the potential, and using the Green's function method for the analysis, they reached the conclusion that only a very deep potential could explain the data. 

\begin{figure}[htb]
 \begin{center}
\includegraphics[width=.5\textwidth]{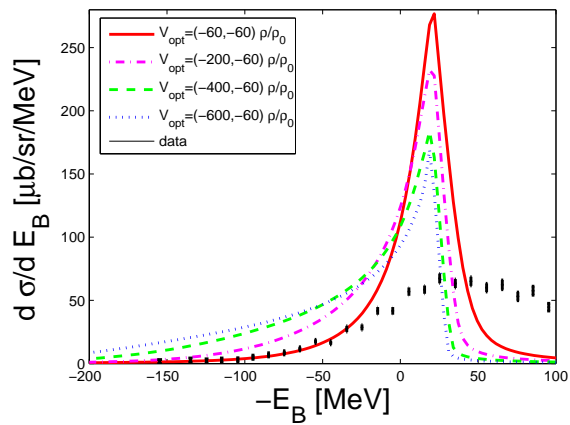}
\caption{Results obtained using the many body method for kaon potential depths
of $60$ MeV, $200$ MeV, $400$ MeV and $600$ MeV at normal nuclear density. 
Experimental data are shown with black bars.}
\label{fig23a}
\end{center}
\end{figure}

  In a recent work a theoretical study of the reaction has been done 
   \cite{Magas:2009dk} including other reaction channels than the only one considered in \cite{Kishimoto:2007zz}, the elastic $K^- N \to K^- N$ reaction.
   As one can see in fig. \ref{fig23a}, we find a cross section for the reaction much bigger that experiment independently of the depth of the potential assumed.  
   
\begin{figure}[htb]
\begin{center}
\includegraphics[width=0.6\textwidth]{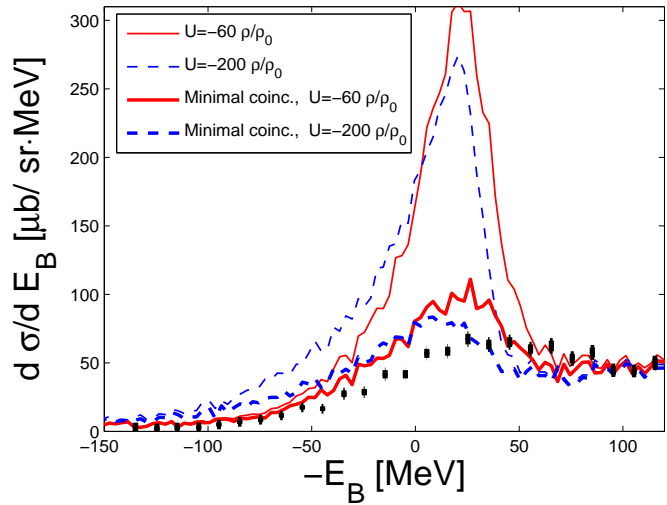}\\
\caption{ Proton spectra without (thin lines) and with (thick lines)
the minimal
coincidence requirement, using
$V_{\rm opt}=(-60,-60)\rho/\rho_0$ MeV (solid lines) and  
$V_{\rm opt}=(-200,-60)\rho/\rho_0$ MeV
 (dashed lines). Experimental data are 
  from \cite{Kishimoto:2007zz}.}
\label{dekishi}
\end{center}
 \end{figure}
   
 The way out of the puzzle came from the realization of a subtle detail in the implementation of the experiment. Indeed, a seemingly innocent sentence in the experimental paper reads " ...we demand that there is at least one charged particle detected in coincidence with the fast proton in a layer detector surrounding the target."  In the experimental paper it was assumed that this coincidence requirement did not affect the shape of the distribution. Given the geometry of the target and the layer, it is easy to see that in cases where the recoiling $K^-$ goes backward and does not interact with other particles in the nucleus, then there will be no charged particle detected in this layer detector. We call this a minimal coincidence requirement in our simulation of the experiment, the results of which are shown in Fig. \ref{dekishi}. What we see in this figure is that the imposition of this minimum requirement reduces the cross section by about a factor of three and the shape is drastically changed with respect to the one with no coincidence demanded.

 One should expect that in some other cases, where there is a $K^-$ collision, there are no charged particles produced, such that our results should overestimate the experiment, as it is indeed the case, but the minimal coincidence requirement has brought us much closer to the experiment. One can also see  in the figure that taking a depth of 60 MeV or 200 MeV does not change much the cross section in the region corresponding to kaons experiencing an attraction by about 50 MeV or more. The main finding of the work of \cite{Magas:2009dk} is that the reaction is not too suited to determine the depth of the kaon nucleus optical potential, invalidating the conclusions drawn in the experimental work of \cite{Kishimoto:2007zz}, since the theoretical analysis of the spectrum in \cite{Kishimoto:2007zz} was done without consideration of the coincidence requirement, which, as we have shown, changes drastically the cross section and its shape with respect to the real spectrum. 
 
    Very interesting work on the possible bound $K^-pp$ clusters is an on-going one which has been reported in this session by Wolfram Weise. Once again, the theoretical results differ from each other, some of them giving large binding energies of the order of 50-70 MeV \cite{Shevchenko:2007zz}, while others, based on the underlying chiral dynamics, produce more moderate bindings, of the order of 20 MeV, and much larger widths, of the order of 50-70 MeV  \cite{Dote:2008in}. An important step in the clarification of this issue has been done by the recent work of \cite{Ikeda:2010tk}, where it is shown that the energy dependence of the kernel of the interaction (potential) is the main ingredient in giving different theoretical results, being the energy dependence provided by the chiral Lagrangians the one leading to the smallest binding energies. 
 
\section{Charmed mesons in a nuclear medium}
    Charmed mesons in a nuclear medium is a hot topic since its prospects of study at the forthcoming FAIR facility have stimulated work in this direction.  There is work done on the interaction of $D$ and $\bar{D}$ in a nuclear medium 
\cite{Lutz:2005vx,Mizutani:2006vq,Tolos:2007vh}, and even studies of possible 
$D$ bound mesic nuclei \cite{GarciaRecio:2010vt}. In this section, however, we wish to present recent results on the renormalization of meson properties in matter for some mesons which are dynamically generated with the dynamics of the hidden gauge Lagrangians  \cite{Bando:1987br}. In \cite{Kolomeitsev:2003ac,Guo:2006fu,Gamermann:2006nm} the interaction of mesons with charm with other mesons is studied and a few states, which can be associated to recently observed resonances with open charm, are obtained as dynamically generated resonances. One of them, well known in the literature, is the $D_{s0}(2317)$ which mainly couples to the $DK$ system. In \cite{Gamermann:2006nm} the hidden charm states are also investigated and one scalar isoscalar state is found around 3700 MeV which couples mostly to $D\bar{D}$. This latter state is not officially in the PDG but might have been observed by the Belle collaboration \cite{Abe:2007sya} according to the reanalysis of Ref.~\cite{Gamermann:2007mu}. In this section we wish to report on the study done in \cite{Molina:2008nh} on the modification of the mass and width of these two resonances in the nuclear medium.

The $D_{s0}(2317)$ and X(3700) resonances are generated dynamically solving the coupled-channel Bethe-Salpeter equation for two pseudoscalars \cite{Molina:2008nh}. The kernel is derived from a $SU(4)$ extension of the $SU(3)$ chiral Lagrangian used to generate scalar resonances in the light sector. The $SU(4)$ symmetry is, however, strongly 
 broken, mostly due to the explicit consideration of the masses of the vector 
 mesons exchanged between pseudoscalars \cite{Gamermann:2006nm}. 

The transition amplitude of each resonance to the different coupled channels gives us information about the coupling of this state to a particular channel. The $D_{s0}(2317)$ mainly couples to the $DK$ system, while the hidden charm state $X(3700)$ couples most strongly to $D\bar{D}$. Thus, while the $K$ and $\bar{D}$ self-energies are small compared to their mass, any change in the $D$ meson properties in nuclear matter will have an important effect on these  resonances. Those modifications are given by the $D$ meson self-energy in the $SU(4)$ model without the phenomenological isoscalar-scalar term, but supplemented by the $p$-wave self-energy through the corresponding $Y_cN^{-1}$ excitations \cite{Molina:2008nh}.

\begin{figure}
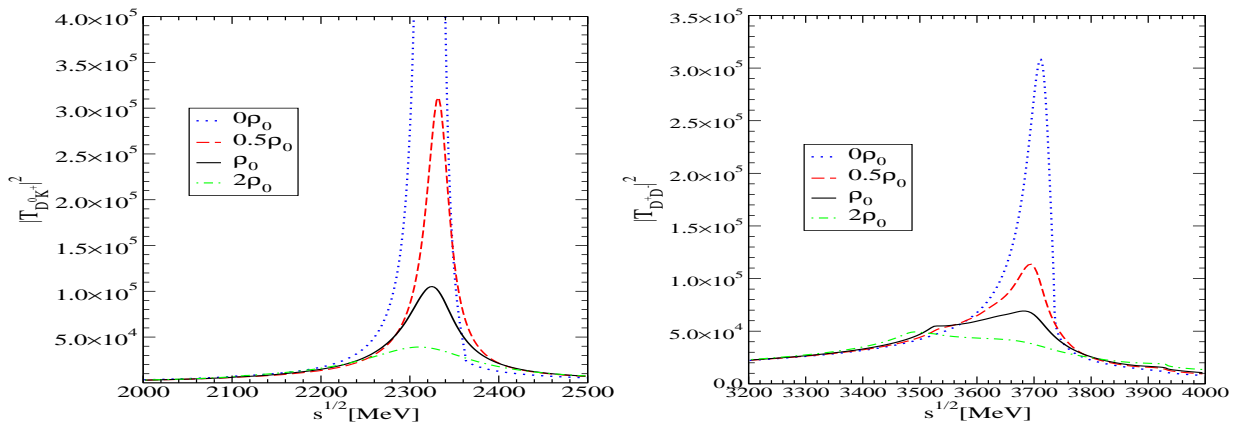

\includegraphics[width=0.5\textwidth,height=5.5cm]{ds02317}
\hfill
\includegraphics[width=0.5\textwidth,height=5.5cm]{x37}
\caption{$D_{s0}(2317)$ (left) and  $X(3700)$ (right) resonances in nuclear matter. \label{fig2}}
\end{figure}

 In Fig.~\ref{fig2} the $D_{s0}(2317)$ and $X(3700)$ resonances are displayed via the squared transition amplitude for the corresponding dominant channel at different nuclear densities. The $D_{s0}(2317)$ and $X(3700)$ resonances, which have zero and small width,
develop widths of the order of 100 and 200 MeV at normal nuclear matter density,  respectively. This is due to the opening of new many-body decay channels as the $D$ meson gets absorbed in the nuclear medium via $DN$ and $DNN$ inelastic reactions. We do not extract any clear conclusion for the mass shift. We suggest to look at transparency ratios to investigate those in-medium widths. As we have discussed before, this magnitude, which gives the survival probability in production reactions in  nuclei, is very sensitive to the in-medium width  of the resonance \cite{Kaskulov:2006zc,Hernandez:1992rv}.

\section{Nuclei with many mesons instead of baryons?}

    One may wonder why the known nuclei are made of baryons and not of mesons. Certainly, it is not that the interaction between mesons is weaker than between baryons. The reason lies in the property of baryon number conservation which does not hold for mesons. As a consequence, an aggregate of protons and neutrons sufficiently bound has nowhere to decay if baryon number is conserved. However, and aggregate of mesons would decay into systems of smaller number of mesons. Let us then accept that these systems will be unstable, but, even then, could we see them as resonances with a certain width, as most of the particles in the PDG?. Gradually an answer is coming to this question. Bound states or resonances of two mesons have become common place thanks in part to developments in chiral unitary theory \cite{Oller:2000ma,Kaiser:1998fi}. One step further in this direction  was given in the study of the three body system $\phi K \bar{K}$ which leads to the recently discovered state X(2175), as has been shown in \cite{MartinezTorres:2008gy}. This system, studied through Faddeev equations in coupled channels, produces a state in which the $K \bar{K}$ pair clusterizes into the $f_0(980)$ resonance and the $\phi$ interacts with this cluster. One could proceed further and investigate more complex meson systems. This is what has been recently done in \cite{Roca:2010tf} where the $f_2(1270)$, $\rho_3(1690)$,   $f_4(2050)$,
 $\rho_5(2350)$ and   $f_6(2510)$ resonances have been described as multi-$\rho(770)$ states.
 
  The idea behind the work of \cite{Roca:2010tf} is that recent studies show that the $\rho \rho$ interaction is very strong \cite{Molina:2008jw,Geng:2008gx}, particularly when the two $\rho$ mesons align their spins to form a state of spin S=2.
  This interaction is so strong that can bind the two $\rho$ mesons leading to a bound state which, according to \cite{Molina:2008jw,Geng:2008gx}, is the $f_2(1270)$.
 It is surprising to come out with this idea when it has been given for granted that this state and other partners accommodate easily as $q \bar{q}$ states and can reproduce most of the known properties of these states 
\cite{Fariborz:2006xq,Umekawa:2004js,Anisovich:2001zp}. However, it has been shown that with this molecular picture one can reproduce the radiative decay into $\gamma \gamma$ \cite{junko}, the decay of $J/\Psi$
into $\omega (\phi)$ and $f_2(1270)$ (together with other resonances
generated in \cite{Geng:2008gx})  \cite{daizou}, and $J/\Psi$ into $\gamma$
and  $f_2(1270)$ (and the other resonances of \cite{Geng:2008gx})
\cite{hanhart}.  

   Once this is accepted, the idea is to study  a system with three $\rho$ mesons. To give the maximum probability of binding we choose them with their spins aligned to give a state of S=3.  The interaction is studied using the fixed center approximation (FCA) to the Faddeev equations. One obtains the scattering matrix and looks for peaks in $|T|^2$, from where one obtains the mass and the width of the states. One finds a peak that we associate to the  $\rho_3(1690)$. Once this is done then one takes two clusters of $f_2(1270)$ and studies their interaction using as input the scattering matrix $\rho ~ f_2(1270)$ obtained before. The peak obtained in $|T|^2$ can be associated to the    $f_4(2050)$. One further step uses the FCA to study the interaction of a $\rho$ with the $f_4(2050)$, using as input the $\rho ~ f_2(1270)$ interaction obtained before. In this way one obtains a peak that is associated to the  
   $\rho_5(2350)$. A further step, letting an $f_2(1270)$ interact with the 
  $f_4(2050)$ previously obtained, is done and then a peak in  $|T|^2$ appears which we associate to the  $f_6(2510)$.

\begin{figure}
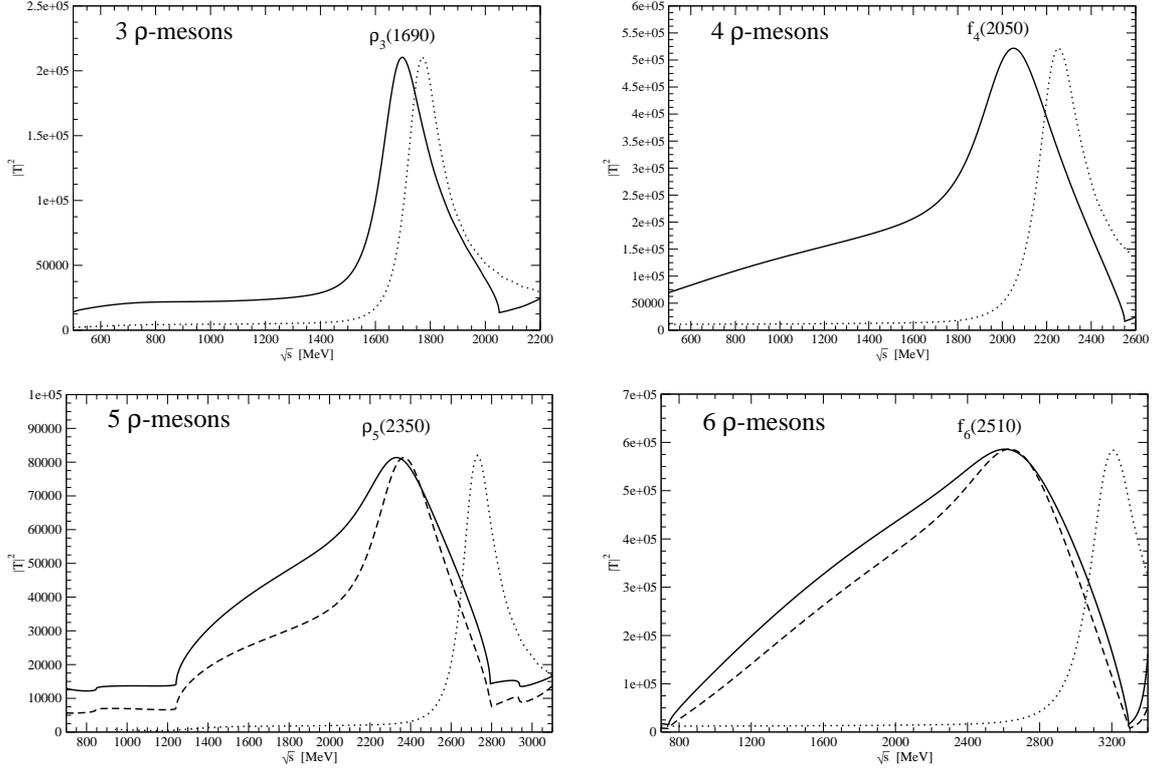

\begin{center}
\makebox[0pt]{\includegraphics[width=.45\linewidth]{figure5a.eps}\hspace{0.5cm}
\includegraphics[width=.45\linewidth]{figure5b.eps}}\\\vspace{0.3cm}
\makebox[0pt]{\includegraphics[width=.45\linewidth]{figure5c.eps}\hspace{0.5cm}
\includegraphics[width=.45\linewidth]{figure5d.eps}}
   \caption{Modulus squared of the unitarized multi-$\rho$ amplitudes.
   dotted line: only single-scattering. Solid lines correspond to the prediction of the model. Dashed lines come from making a small change in a cut off. The solid one is the one used. 
 (The dashed and dotted lines have been normalized to
   the peak of the solid line for the sake of comparison 
   of the position 
   of the maxima)}
     \label{fig:T2s}
\end{center}
\end{figure}

In fig.~\ref{fig:T2s} we show the modulus squared of the amplitudes for
different number of $\rho$ mesons considering only the single scattering
mechanisms (dotted line) and the full model (solid and dashed lines).
The difference between the solid and dashed lines is the value of 
$\Lambda'\big|_{f_4}$ (see  \cite{Roca:2010tf})
needed in the evaluation of the
$5\rho$ and $6\rho$ meson systems
(1500~MeV 
in the
solid line, 875~MeV in the dashed one).
The
dotted and dashed curves have been normalized to the peaks of the
corresponding full result for the sake of comparison of the position of
the maximum.
The difference between the dashed and solid lines can be
considered as an
estimate of the error but the variation in the position of the maximum
is small.

We clearly see that the amplitudes show  
pronounced bumps which we associate to the resonances 
labeled in the figures.
The position of  
the maxima can be associated to the masses of the corresponding
resonances. 

\begin{figure}[!t]
\begin{center}
\includegraphics[width=0.6\textwidth]{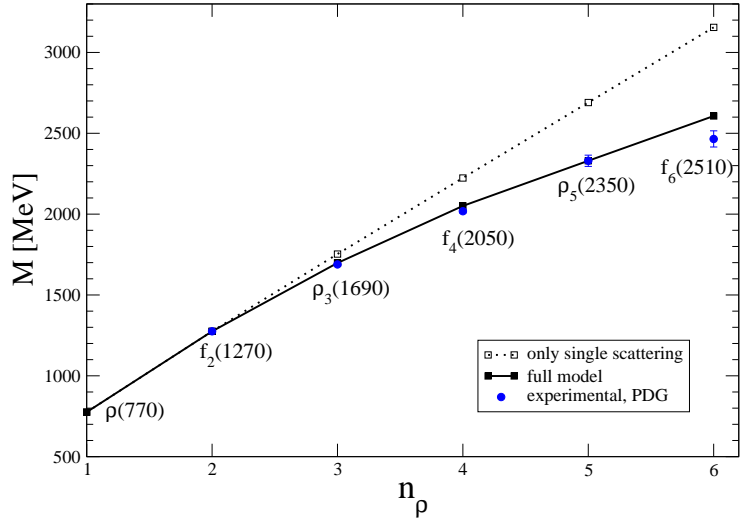}
\caption{Masses of the dynamically generated states as a function of the
number of constituent $\rho(770)$ mesons, $n_\rho$. Only single
scattering contribution (dotted line); full model (solid line);
experimental values from the PDG, (circles).
}
\label{fig:Mvsn}
\end{center}
\end{figure}

 A perspective of the masses obtained compared with those in the PDG can be seen in fig. \ref{fig:Mvsn}. As we can see, the agreement obtained is excellent. Granted that Nature is always more subtle that any picture that we can make of it, the agreement found for the different states is certainly impressive, with no free parameters used. This is certainly a result worth thinking about which should encourage further searches in this direction with likely interesting surprises ahead.

\section{Conclusions}
   As a general summary of the lessons one has learned from all these works we would like to mention a few things: 
\begin{enumerate}

\item The manybody effects are spectacular in some cases:
They teach us about excitation mechanisms of the nucleus,
but also about elementary properties of some resonances which are tied
to their composition.

\item We have seen repeatedly that it is fair and finally rewarding to ask for extra rigor in the experimental analyses.

\item Charm physics is coming into play. FAIR will devote energies to this end and to properties of charm and hidden charm states in nuclei.
 
\item Watch for multihadron states (not ordinary nuclei), they were
always there but only now they are emerging. Facilities and collaborations like  BELLE, COMPASS, WASA/COSY, etc., will have something to say about this topic in the future. 

\end{enumerate}

 \section*{Acknowledgments}
L.T. acknowledges support from the RFF program of the University of Groningen. This work is partly supported by the EU contract No. MRTN-CT-2006-035482 (FLAVIAnet), by the contracts FIS206-03438, FPA2007-65748, FIS2008-01661 and FIS2008-01143 from MICINN (Spain), the Generalitat Valenciana in the program Prometeo, the SA016A07 and GR12 projects of la Junta de Castilla y Leon, by the Spanish Consolider-Ingenio 2010 Programme CPAN (CSD2007-00042), by the Generalitat de Catalunya contract 2009SGR-1289 and by Junta de Andaluc\'{\i}a under contract FQM225. We acknowledge the support of the European Community-Research Infrastructure Integrating Activity ``Study of Strongly Interacting Matter'' (HadronPhysics2, Grant Agreement n. 227431) under the 7th Framework Programme of EU.

\end{document}